\documentclass[prl,nofootinbib]{revtex4}
\usepackage{graphicx}
\usepackage{latexsym}
\def\be{\begin{equation}}
\def\ee{\end{equation}}
\def\bea{\begin{eqnarray}}
\def\eea{\end{eqnarray}}

\begin{document}
\title{Relativistic Spacetime Based on Absolute Background}

\author{Chi-Yi Chen${}^{a}$}
\email{iamchiyi@gmail.com; chenchiyi@hznu.edu.cn}

\affiliation{${}^a$Hangzhou Normal University, Hangzhou 310036, China}

\begin{abstract}
Based on the consideration of naturalness and physical facts in Einstein's theories of relativity, a nontrivial spacetime physical picture, which has a slight difference from the standard one, is introduced by making a further distinction on the absolute background of spacetime and the relative length or duration of base units of spacetime. In this picture, the coordinate base units in gravity-induced spacetime metric are defined by the standard clock and ruler equipped by the observer, and duplicated onto the every position of the whole universe. In contrast, the local intrinsic base units of spacetime in gravitational field are defined by the length and duration of physical events intervals in the same-type standard clock and ruler which are really located at every position of the universe. In principle, the reading number of the standard clock is counted by the undergone times of unit intervals defined depending on a certain kind of local intrinsic events. But the size of the base units of spacetime is essentially depicted by the length of the line segment, which is cut from the absolute background of spacetime by the local intrinsic events of unit interval. The effect of gravitation is just to change the length of this segment for base spacetime units. On the basis of such a physical picture of spacetime, in a fairly natural way we re-derive a new classical dynamical equation which satisfies a more realistic and moderately general principle of relativity. To further examine this physical picture including of gravitation and spacetime, we also reinterpret the gravitational redshifts for solar gravity tests.
PACS number(s): 45.20.-d, 45.05.+x
\end{abstract}

\maketitle
\section{1 INTRODUCTION}
As well known to us, the gravitational redshift experiment is one of three traditional verification tests for Einstein's general theory of relativity. The traditional interpretation of the gravitational redshift effect is based on the spacetime physical picture given by Einstein's equivalence principle\cite{weinberg,Stewart}, which should be distinguished from the weak equivalence principle, namely the equality between the gravitational mass and inertial mass. As a generalization of weak equivalence principle, Einstein's equivalence principle further claims that the gravitational force must be equivalent to the inertial force on all their physical effects\cite{Ohanian}. In the spirit of Einstein's equivalence principle, all free-falling reference frames under gravity are regarded as local inertial reference frames and all of them are equivalent. In other words, the spacetime properties in these local inertial reference frames are assumed to be exactly the same. The standard interpretation of gravitational redshifts from the point of view of general relativity has been discussed in detail in many textbooks\cite{weinberg,Stewart,Ohanian,Liu}. Frankly speaking, whether the inertial force can be equivalent to the gravitational force is worthy of further investigation. Moreover, how Einstein's general principle of relativity can be exactly proved is still a pending question. Therefore, if a new physical picture of spacetime does not conform to that of general theory of relativity on this point, it will not be doomed to be hopeless. In fact, in this paper we just aim to propose a compromise on this issue. The main idea is that the concept of spacetime should be further subdivided into two aspects. One is the relative length or duration of base units of spacetime. And the other is the absolute background of spacetime. The length or duration of base units of spacetime can be regarded as a unit line segment which is cut from the absolute background of spacetime.

Logically speaking, to any object (with finite size), as long as it exists, there must be a background of the existence. Otherwise, it will make no sense for the concept of production, as well as annihilation, transformation and evolution. Any production, evolution, transformation and annihilation of a specific object must occur relative to a background as the reference. Therefore, in physical logic, the background of an object is just the premise of the existence of this object, and the reference basis only on which any change of the object can be observed. As an analogy, the absolute background of spacetime should be defined as the premise of the existence of the relative length or duration of base units of spacetime, and the reference basis only on which any change of the length or duration of the base units of spacetime can be observed.

The necessity of introducing the concept of an absolute background for spacetime can be illustrated by the following picture. We assumed that there are two spatial points exist in a map. For instance, the distance between these two spatial points is defined as one meter (the base unit of a standard ruler). Now owing to a gravitational field, the observer may find that the one meter in this map is not equal to that equipped by the observer who is located outside the gravitational field. In Einstein's general theory of relativity, this phenomenon may be explained by that the space in the map is contracted. But in fact there is another more simple interpretation. It is that the size of the drawing board as the background of whole map is actually not changed. Nevertheless, the spatial span of the defined one meter in the map as the base unit of a standard ruler is changed under the gravitational field. As an analogy, here the invariant drawing board of the map is just equivalent to the absolute background of space, and the distance between two assigned spatial points is just equivalent to the base unit of space. Such an alternative physical picture is superior to the original physical picture in Einstein's theories of relativity. In the original physical picture, the map is contracted spatially. It either should be compensated by a stretching effect of space around the map, or be realized by a global movement toward the contractive center of map. All these physical effects are unnatural.

Based on a proposed fundamental picture for spacetime, this paper also aims to reinvestigate main physical proofs relevant to the properties of spacetime. Sec.1 is an introduction. In Sec.2, the compatibility between the concept of background of spacetime and Einstein's theories of relativity is preliminary discussed. In Sec.3, a more specific physical picture including of the relative length or duration of base units of spacetime and the absolute background of spacetime is presented in detail. In Sec.4, it is shown that a new formalism of particle dynamics can be naturally derived under the framework of classical mechanics, based on above physical picture of spacetime. In the light of the correctness of this new dynamical equation, the existence of an absolute background for spacetime is also strongly supported. In Sec.5, considering the nature of inertial force demonstrated in the new formalism of particle dynamics, the physical picture for changes of the length or duration of base units of spacetime is self-consistently established. In Sec.6, to further examine the physical picture for changes of the length or duration of base units of spacetime, we reinterpret the gravitational redshift effect by retaining Einstein's gravitational field equation. Finally in Sec.7, the verifiability of proposed physical picture of spacetime is discussed.

\section{2 COMPATIBILITY BETWEEN EINSTEIN'S THEORY OF RELATIVITY AND PHYSICAL CONCEPT OF BACKGROUND OF SPACETIME}
The currently admitted physical theories of spacetime are Einstein's special theory of relativity and general theory of relativity. But if we reinvestigate the main physical logic in these two theories, we will find that both of them can substantially be understood as the change rules of the length or duration of base units of spacetime\cite{Sanhui,weinberg}. Just as its name implies, the length of base unit of space is the spatial span of the standard one meter, and the duration of base unit of time is the time span of the standard one second. In physical pictures of Einstein's theories of relativity, the concept of background of spacetime is not deliberately distinguished from the concept of the base units of spacetime. There is no concept of the background of spacetime in Einstein's theory. But in fact, the physics of the background of spacetime has been implicitly included in both Einstein's theories of relativity.

1), in the physical logic of special theory of relativity, every event is assumed to have an objective position in spacetime manifold when its coordinates are transformed between arbitrary two inertial reference frames. Otherwise, the Lorentz coordinate transformation cannot be obtained. Here the objective position of a physical event in spacetime manifold means that the event's occurring point in spacetime manifold doesn't change with the inertial reference frames. The existence of an objective position in spacetime manifold can actually be regarded as the reflection of the existence of an absolute background of spacetime\cite{Sanhui}.

2), in the physical picture of general theory of relativity, gravitational fields will result in a dilation effect for the duration of base unit of clocks and a contraction effect for the length of base unit of rulers. In other words, the span of base units defined in standard clock or standard ruler will be changed owing to the existence of a gravitational field. But theoretically, we should have a deeper picture for these physical effects. How can such a change in the spatial span of the base unit in standard rulers be embodied? There may be only one answer survivable. That is the existence of an absolute background of spacetime. Only when the base units of spacetime are compared with the absolute background of spacetime, the changes in their length or duration can be reflected and such an effect can be physical. More specifically, the base unit of standard ruler or stand clock is directly defined by the unit interval between two local intrinsic physical events which periodically occurs in specific objects, and the spatial span of the base unit of the standard ruler is just the line segment which is cut from the background of spacetime by the corresponding two local intrinsic events of the given unit interval.

Therefore at least in physical concepts, the existence of the background of spacetime can be compatible with the length or duration of base units of spacetime in a generalized physical picture based on Einstein's theories of relativity. Moreover, it is necessary to distinguish these two concepts, since the length or duration of base units of spacetime are relatively changeable according to Einstein's theory, but the background of spacetime must be absolute. Regarding the absolute background of spacetime, Natan Rosen has ever proposed a kind of bi-metric theories. He introduced an extra metric for flat space, in parallel to Einstein's curved metric, and both of them coexist in his theory\cite{Rosen}. The concept of flat space in Rosen's theory is a little close to here absolute background of spacetime, but they are different. Viewed from the side, Rosen's flat space is at least induced from his bi-metric theory to modify the general theory of relativity. But in our theory, there is always only one metric get involved. The spacetime metric normally describes the curve of spacetime or the change of length and duration of base units of spacetime under the existence of gravitation. And if all the matter in the universe is entirely absent, the spacetime metric must be reduced to be that of a flat Minkowski spacetime which substantially describes the absolute background of spacetime with the mathematically introduced base units of spacetime by observers.

\section{3 RELATIVE BASE UNITS AND ABSOLUTE BACKGROUND IN THE PHYSICAL PICTURE OF SPACETIME}
\subsection{3.1 Physical concepts}
First of all, the length or duration of base units of spacetime and the background of spacetime are essentially the two aspects of spacetime, instead of two kinds of spacetimes. In physical concept, the absolute background of spacetime should be defined as the premise of the existence of relative length or duration of base units of spacetime and the reference basis only on which any change of the length or duration of base units of spacetime can be observed. In essence, the length or duration of base units of spacetime can be regarded as a unit line segment which is cut from the absolute background of spacetime. Taking a flat two-dimension plane for example, only after we define the length or duration of coordinate base units, a coordinate system is able to be painted on, and so we have a measurable concept of length for spatial spans on this two-dimension plane. But how far is the length of one meter of the standard ruler? To answer this question, the bottom board of this plane is indispensable. Without this bottom board serving as a foil to reflect, the length of one meter for the standard ruler will not make any sense. As an analogy, the background of spacetime is just equivalent to the bottom board of this two-dimension plane. If we ponder over it more deeply, the four-dimension background of spacetime may be imagined as a blank sheet of four-dimension paper. Originally, there is no coordinate on it. It is nothing but the observation that requires the introduction of coordinate base units. We can only define the base coordinate units by resorting to the local intrinsic events which periodically occur in specific objects, so the coordinate system is established.

Secondly, the base units of spacetime are physically defined by the unit intervals of local intrinsic events which occur in specific objects, on account of the requirement of measurements from observers. For instance, the second is the base unit of time in the International System of Units (SI). Since 1967, the second has been defined to be the duration of 9192631770 periods of the radiation corresponding to the transition between the two hyperfine levels of the ground state of the Caesium 133 atom\cite{BIPM}. Therefore in principle, the interval of spacetime between local intrinsic events is able to change, but the background of spacetime as the basis to reflect this change so must be homogeneous forever.

\subsection{3.2 The length or duration of the base units of spacetime}
Essentially, the base units of spacetime are directly defined by the unit intervals of local intrinsic events which periodically occur in specific objects. In this way, the length or duration of a spacetime base unit is described by the span of the line segment which is cut from the background of spacetime by the corresponding two local intrinsic events. There are long and short line segments. There are thus large and small base units of spacetime. Specifically, the duration of base unit of time can be denoted by the length of the line segment (or duration): $\overline{\Delta\tau(=1)}$, which is cut from the background of spacetime by the unit interval of local intrinsic events (${\Delta\tau=1}$).

\subsection{3.3 The reading number of the ruler and clock}
In contrast, the reading number of observers' clocks and rulers are substantially determined by the number of times for local intrinsic events which occur. Therefore, the reading number of clocks or rulers itself does not directly contain any information of spacetime base units. Since every physical event has its objective position in the background of spacetime, the change of a spacetime base unit can be determined by making a comparison of the fore-and-aft reading numbers of clocks or rulers as long as their corresponding line segments have the same length in the background of spacetime. The reading number of local intrinsic clock is just the time recorded by local observer, which can be denoted by $\Delta\tau$. Consequently, it should definitely be able to distinguish the reading number of local intrinsic clocks ($\Delta\tau$) and the length of the corresponding line segment cut from the background of time ($\overline{\Delta\tau}$), for every interval of events. In physical concepts, we should distinguish the local intrinsic clock and the observer's clock. The interval $dt$ or $dr$ appears in the curved invariant spacetime interval ($ds$) is usually the reading number of the observer's clock or observer's ruler.

\subsection{3.4 The changeability of the length or duration of physically defined base units of spacetime}
In special theory of relativity, a relative velocity between different reference frames will result in a dilation effect for the duration of clocks and a contraction effect for the length of rulers. It means that a spacetime base unit which is originally defined by the unit interval of the same kind of local intrinsic events may be different in length in the eyes of different observers. Nevertheless, in special theory of relativity both effects are simultaneously valid for each other of two observers, so it can be understood as an observational effect\cite{Sanhui}.

In general theory of relativity, a gravitational field will also result in a dilation effect for duration of clocks and a contraction effect for the length of rulers. Therefore, a spacetime base unit which is originally defined by the unit interval of the same kind of local intrinsic events will change in length or duration under the different gravitational field strength. In other words, there is a relative evolution which may exist for a local intrinsic clock. According to solar gravity tests, the geometric theory of gravity can be understood as a physical law for the change of the length or duration of physically defined base units of spacetime. As for the relation between the local intrinsic base units of spacetime and the observer's base units of spacetime, there is a really important assumption introduced. It is that the gravity causes the curvature of spacetime but in an infinitesimal neighborhood the spacetime should be asymptotically flat.

\subsection{3.5 The flatness, homogeneity and absoluteness for the background of spacetime}
First of all, it should be pointed out that any concept of flatness, homogeneity and absoluteness for any object should be defined by comparing with a more basic reference background. Therefore, if the background for the spacetime in whole universe has been set to be the most basic background, in the eyes of the observer, it should congenitally be regarded (or defined) to be flat and homogeneous. Because once it is not flat or homogenous, then such a conclusion must be made based on a more basic reference object. But as what we have just defined, the background for the spacetime in whole universe is set as the background at the most fundamental level. Therefore, it is always valid to say that the background of spacetime is flat and homogeneous. Similarly, we can always say that the background for the spacetime in whole universe is absolute. The reason is that we have defined the background for the spacetime in whole universe as the remained physical state after we have removed away all movable or evolvable objects from the current universe. Therefore, once the background is not absolute, it implies that this background must evolve with respect to a more basic reference object. However we have set the background for the spacetime in whole universe as the most fundamental reference. Consequently, it is also valid to say that the background for the spacetime in whole universe is absolute.
¡¡
\subsection{3.6 The preexistence and perpetuity for the background of spacetime}
There are many discussions about the creation of the universe in modern cosmology\cite{Grishchuk,Ferrara}. But incorporating above physical picture of spacetime, there one point which must be made clear is that, the so-called creation of the universe should be only limited to matter in our observable universe, instead of the background for the spacetime in whole universe. If the whole universe is really created from a thorough nothing, it means that such a creation doesn't require any premise or any precondition. Therefore, new universes would be created anytime and anywhere. This is not true. In this sense, the background of spacetime should preexist and last forever.

Besides, it is also meaningful to discuss the simultaneity in the background of time. In principle, the simultaneity in the background of time always exists according to a basic hypothesis that the background of time passes homogeneously. But an observable simultaneity should be artificially defined. For instance, if an observer wants to make clear the simultaneity between different spatial positions by means of the observation of physical phenomena, he has to resort to the number of times of local intrinsic events which occur on these spatial positions. In other words, the observable simultaneity should be determined by the coordinate values of spacetime manifold. Moreover, if we want to make a precise definition of the observable simultaneity, some physical interaction with invariant propagation speed may be required. For example in Einstein's special theory of relativity, this observable simultaneity is defined by the principle of the invariance of light speed, which is placed top priority. Therefore, an observable simultaneity is not always available for us in many cases. But for two events which occur on the same spatial position, we will definitely be able to distinguish the time order of the occurrence, so we always can retain the concept of simultaneity for the same spatial position. Therefore, the simultaneity in the background of time always exists objectively. But the directly observable simultaneity for observers must be defined by resorting to specific physical phenomena.

\section{4 ABSOLUTE BACKGROUND AND NEW FORMALISM FOR CLASSICAL PARTICLE DYNAMICS}
In the framework of Newtonian mechanics, the fundamental dynamics equation is Newton's second law.  But as is well known, Newton's second law is only valid in inertial reference frames. Provided that we apply the same equation of Newton's second law in a non-inertial reference frame, we need to introduce a fictitious force---inertial force additionally. The magnitude of the inertial force is usually determined by the relative acceleration between the non-inertial reference frame in question and a certain inertial reference frame\cite{Shouzhu,DOUGLAS}. Therefore, the Newtonian particle dynamics is totally based on the concept of inertial reference frame. However, we are never able to find a real inertial reference frame in practice. This situation is surely not satisfactory\cite{ESSEN,Liu}.

On the other hand, the particle dynamical law which is applied very successful in practice and deeply accepted by people is actually empirical laws. The empirical laws are not totally equivalent to the theoretical formula of Newton's second law. The reason is that a theoretical formula of Newton's second law is only valid in inertial reference frames. But all real reference frames used in practice are not exactly the inertial reference frame. Besides, in empirical laws the term of exerted forces does not need to take into account the total force acting on the particle. Understanding this subtle difference is the key point to understand the physical meaning of the following reformulated particle dynamics.

But above all, there actually is a problem of causal inconsistency and dissymmetry which exists in the theoretical formula of Newton's second law. In principle, Newton's second law should be a causal law of particle dynamics. Here the forces acting on the particle under study should be the cause and the resulting acceleration should be the effect. In history, huge amounts of experiments of classical mechanics had also illustrated a quasi-differential causal relationship between the new additionally exerted force (compared with a previous mechanical state) and the resulting relative acceleration under the premise of reference frame being fixed: $\Delta{\bf F}=m\Delta{\bf a}$. However, the traditional formula of Newton's second law is given by
\begin {eqnarray}
{\bf F}|_{p}=m_{p}{\bf a}|_{p-O}.
\end {eqnarray}
In theory, the left hand side of this equation (${\bf F}|_{p}$) must denote the total forces from the whole universe acting on the particle $p$. Otherwise, when the equation is applied into practical cases, we will not be able to make it clear what forces should be taken into account, and what forces should not be taken into account. The left hand side (${\bf F}|_{p}$) only depends on $p$. Yet the right hand side ${\bf a}|_{p-O}$ is the acceleration of the particle $p$ with respect to the reference frame $O$, equivalently measured relative to the reference object of $O$ which corresponds to the origin point of the reference frame. Therefore in fact, the effect (namely the result) ${\bf a}|_{p-O}$  depends not only on the particle $p$, but also on the reference object of $O$. In this sense, the causality of Newton's second law is not symmetric and consistent. This is the very point to account for why Newton's second law is theoretically valid only in so-called inertial reference frames, but none of them can be found in practice.

Since for Newton's second law, neither the theory nor the causality is satisfactory, we consider whether it is possible to reconstruct the physical logic for particle dynamics. Firstly, we accept the empirical laws summarized from a huge number of classical mechanics experiments, namely the quasi-differential causal relationship between the new additionally exerted force  and the resulting relative acceleration under the premise of reference frame being fixed. This causal relationship can be depicted by $\Delta{\bf F}=m\Delta{\bf a}$. Secondly, we accept above proposed physical picture of spacetime which distinguishes concepts between the relative length or duration of physically defined base units of spacetime and the absolute background of spacetime. On this basis, we start to explore a new formalism of particle dynamics using logical deduction.

In this process, the only one most fundamental principle which can be resorted to is the causal consistency principle. Since the particle dynamics is certainly to be a theory with causal principle, it is natural to regard forces as the cause, and regard accelerations as the effect. According to the classical mechanics experiments, a differential causal relationship should be given by $d{\bf F}=md{\bf a}$. Then how to solve the problem of causal inconsistency in its integral form? The key point is how to describe the corresponding effect according to the causal consistency principle when the total force from the whole universe acting on the particle is the cause under the consideration. Under the framework of classical mechanics, the total force acting on a single particle should be objective, namely it will not change with the variation of frames of reference. Therefore, the corresponding effect should also be objective, and not relevant to any reference frame. In this way, a completely objective acceleration can only be expressed as the acceleration with respect to the absolute spatial background of the whole universe,
\begin {eqnarray}
{\bf F}|_{p}= m_{p}\frac{d^{2}}{dt^{2}}\bigodot|_{p}.
\end {eqnarray}
Here the objective position of the particle $p$ in the absolute background of space is particularly denoted by $\bigodot|_{p}$. Although every particle has its objective position in the absolute background of space, there is still a problem that the objective position in the absolute background of space cannot be directly measured. What we can really measure is the difference between any two objective positions, which substantially constructs a mathematical vector,
\begin {eqnarray}
{\bf r}|_{p-O}= \bigodot|_{p}- \bigodot|_{O}.
\end {eqnarray}
After that, we are able to construct a particle dynamical equation which is really available to any observers. In fact, every reference frame must be established on a real reference object. Otherwise, there would be no reference value in measuring any object's motion in the natural world. In other words, a physical reference frame must be the real reference frame. All objects in the universe, including objects under study ($p$) and reference objects ($O$), should be of equal status in the most fundamental law of dynamics. For this reason, the dynamics of any real reference object should also satisfy
\begin {eqnarray}
{\bf F}|_{O}=m_{O}\frac{d^{2}}{dt^{2}}\bigodot|_{O}.
\end {eqnarray}
Here the reference object $O$ naturally corresponds to the origin point of a reference frame, so we can establish a reference frame which is irrotational with respect to the absolute background of space. Originally the base units of spacetime which appear in (2) and (4) are mathematically introduced, and the coordinate system is exactly flat and homogeneous. Once the reference object is selected, the base units of the coordinate system can be naturally defined according to the local intrinsic clock and local intrinsic ruler equipped by the reference object. Therefore, up to now, there is only one reference frame gets involved. It is not yet related to the transformation rule between two reference frames. The introduction of reference frames is just to make relative measurements on kinematical quantities. As a causal correspondence, the forces should also be relatively counted in calculation,
\begin {eqnarray}
m_{O}{\bf F}|_{p}-m_{p}{\bf F}|_{O} = m_{p}m_{O}\frac{d^{2}}{dt^{2}}(\bigodot|_{p}- \bigodot|_{O})= m_{p}m_{O}\frac{d^{2}{\bf r}|_{p-O}}{dt^{2}}.
\end {eqnarray}
Finally, we obtain
\begin {eqnarray}
\frac{{\bf F}|_{p}}{m_{p}}-\frac{{\bf F}|_{O}}{m_{O}}={\bf a}|_{p-O}.
\end {eqnarray}
In this equation, the definition of the force and the acceleration are just the same as that in the traditional theoretical formula of Newton's second law (1). ${\bf F}|_{p}$  and ${\bf F}|_{O}$ are the total forces from the whole universe respectively acting on the particle $p$ and the reference object $O$. $m_{p}$ and $m_{O}$ denote the mass of the particle $p$ and the reference object $O$ respectively. In this way, we finally obtain a new formulated particle dynamical equation (6) even under the framework of classical mechanics. The correctness of this new equation including its comparison with empirical laws in classical mechanics has been repeatedly examined\cite{Chen}. The new formalism of particle dynamics (6) is definitely correct under the framework of classical mechanics. But more importantly, the new formalism of particle dynamics has satisfied the requirement of causal consistency, so it presents a more concise physical picture for classical mechanics\cite{Chen}. From the point of view of practice, in the application of the equation (6), the inertial reference frame is no longer required and the inertial force is no longer introduced by hand. For any real reference frame which is irrotational with respect to the absolute background of space, the nature of the inertial force is nothing but the real force acting on the reference object : ${\bf f}_{inertial}|_{O}=-\frac{m_{p}}{m_{O}}{\bf F}|_{O'}$ , and which is supposed to appear in the new dynamical equation (6) according to the principle of causal consistency. To demonstrate the difference between the equation (6) and the theoretical formula of Newton's second law (1), we may rewrite (6) to be,
\begin {eqnarray}
 {\bf F}|_{p}-\frac{m_{p}}{m_{O}}{\bf F}|_{O}=m_{p}{\bf a}|_{p-O}.
\end {eqnarray}
Here the left hand side of this equation can be called as a relative counting of forces. Obviously, the equation (7) has a net term ($-\frac{m_{p}}{m_{O}}{\bf F}|_{O}$) more than Newton's second law, while the other terms are identical. Although this net term is explained as the inertial force, but it essentially is not just a mathematical modification for accuracy. The existence of this net term, as an independent physical correspondence, has strongly suggested the existence of an absolute background for the spacetime of whole universe.

\section{5 THE PHYSICAL PICTURE FOR THE CHANGE OF BASE UNITS OF SPACETIME}
Incorporating the physical facts in Einstein's geometric theory of gravity\cite{weinberg,Stewart,Liu} and new proposed physical picture of spacetime\cite{Chen}, the local base units of spacetime which are physically defined by local intrinsic events must be different from place to place in the gravitational field. In contrast, the coordinate base units are uniformly defined by the only observer's local intrinsic events and homogenously duplicated all over the background of spacetime\cite{Bailey,Hay,tower,aircraft,rocket}, so the difference between the local base units and coordinate base units constitutes the curve of spacetime in gravitational field. In this spirit, the physical picture describing gravitation into a curved metric of spacetime should be given as follows.

Firstly, a reference object and corresponding irrotational reference frame (with respect to the absolute background of the universe) should be selected, so what gravitational forces should be taken into account can be determined according to the principle of causal consistency. Secondly, the coordinate base units of spacetime should be defined according to the standard clock and standard ruler equipped by the observer himself. Thirdly, the coordinate base units of spacetime as the mathematical clock and mathematical ruler which are imaginatively duplicated onto the every position of the background of spacetime. In this way, a rigid and homogeneous spacetime coordinates system is established. Finally, based on this coordinates system, we determine the curvature of spacetime under gravitational fields by making a comparison at every position between the local intrinsic clock (or ruler) and the observer-defined mathematical clock (or ruler).

Besides, the acceleration's physical effect on the real length or duration of physically defined base units of spacetime should be reexamined, since the nature of inertial forces has been interpreted in the new formalism of particle dynamics (6). As we all know, the problem of inertial reference frames and inertial forces are originated from the theoretical structure of Newtonian mechanics. In Einstein's strong equivalence principle, the physical effect of inertial forces is assumed to be equivalent to that of gravitation\cite{equivalence}. But now, the nature of the inertial force is thoroughly interpreted under the framework of classical mechanics. Moreover, the demonstrated nature of inertial forces is surely different with Einstein's assumption\cite{weinberg,Stewart,equivalence}. Therefore, whether the gravitational force and the inertial force are fully equivalent is worth reexamining, especially for their physical effects on the real length or duration of physically defined base units of spacetime. Firstly, the new particle dynamical equation (6) shows a moderately general principle of relativity, which is obviously different with Einstein's view\cite{Liu,Einstein} on the principle of relativity. The nature of the inertial force is the real force acting on the reference object. Hence the so-called inertial force can actually be all kinds of common forces such as friction force, traction force, gravitational force and so on. But so far we know that only the gravitational force has the time dilation effect. Therefore, Einstein's equivalence principle is neither indispensable nor desirable for the realization of a moderately general principle of relativity. Secondly, for the clock which is relatively at rest in the gravitational field and the clock which is free falling in the same place of gravitational field, they differ only in a non-gravitational force and the resulting acceleration. If there really is no gravitational time dilation effect for the free falling clock, it must imply that a non-gravitational force and the resulting acceleration are also able to bring about a time contraction effect for clocks. However, by now there is no such a sign which has been observed and verified in all past experiments. Thirdly, whether the redshift effect can be aroused by the acceleration in principle can be tested in a ground-based laboratory, and there has been some high energy experiments showing that the proper longevity of negative muon is not related to its acceleration\cite{Bailey,Farley}. Therefore, a free-falling local intrinsic clock in a gravitational field will also change its clock rate which depends on the field strength of gravitation. It is reasonable to retain the numerical equality between the inertial mass and gravitational mass since it has a solid foundation from experiments. But the assumption that all free falling clocks in gravitational field still run in a uniform rate should be given up. It is surly not in conflict with practical experience. We may further imagine that if the running rates of all clocks inside a local region slow down at the same rate, the dynamical law inside of this region must also keep invariant and Einstein's weak equivalence principle is still not violated.

\section{6 REINTERPRETATION OF GRAVITATIONAL TIME DILATION EFFECT IN SOLAR GRAVITY TESTS}
On the one hand, the general covariance should be abandoned in principle, since Einstein's (strong) equivalence principle has been given up. Instead, a moderately general principle of relativity which means that the equation is invariant in any irrotational reference frames (with respect to the absolute background of the universe) is proposed for particle dynamics. It is easy to find that the derivation of Schwarzschild metric also actually conforms to this principle of relativity.

On the other hand, although the assumption that all free falling clocks in gravitational fields run in a uniform rate has been given up, the weak equivalence principle should still be retained. The weak equivalence principle is logically enough to account for the existence of a spacetime geometric description for gravitation. Therefore the idea of gravitation being described by a geometric theory still stands up. The mathematical formula of Einstein's gravitational field equation as a solution guessed according to the weak equivalence principle can also be retained\cite{weinberg,Ohanian}. Furthermore, the process to solve for the Schwarzschild metric is irrelevant to Einstein's assumption on the rate of free falling clocks. Consequently, the mathematical form of Schwarzschild metric is finally retained. But now we will use the above new physical picture of spacetime to reinterpret the gravitational redshift effect in solar gravity tests\cite{tower,aircraft,rocket}.

Provided that the observer is located at the infinity from the sun, the full expression of Schwarzschild metric can be written down as,
\begin {eqnarray}
ds^2=-(1-\frac{2GM}{r})dt^2+(1-\frac{2GM}{r})^{-1}dr^{2}+r^{2}d\theta^{2}+r^{2}sin^{2}\theta d\phi^{2}.
\end {eqnarray}
In above equation, the coordinate base unit of spacetime are mathematically defined according to the standard clock and standard ruler equipped by the observer. In other words, the clock and ruler of the observer are duplicated onto the every position of the whole solar system. After that, the time dilation effect is reflected by the difference of the magnitudes between the reading number of local intrinsic clocks ($(1-\frac{2GM}{r})^{\frac{1}{2}}dt$) and the reading number of the mathematically duplicated observer's clock ($dt$) within the same line segment ($\overline{dt}$) cut from the background of time. For instance, at the surface of the sun, we have $(1-\frac{2GM}{r})^{\frac{1}{2}}<1$. Therefore, under the same duration of line segments in the background of time ($\overline{dt}$), the reading number of the local intrinsic clock at the surface of the sun, will be smaller than that of the local intrinsic clock at infinity. In other words, the clock located at the surface of the sun runs slower than that at infinity.

The coordinate time $t$ in the formula (8) is just measured by the mathematical clock of the observer at infinity, which should be initially introduced before the gravity being quantitatively converted into a curved spacetime. Here the mathematical clock is defined to run at a rigid and homogeneous rate. Therefore, the coordinate time $t$ can be regarded to be equivalently measured by a mathematical background clock. For two events which occur on the same spatial coordinate point, the time intervals can be respectively measured by the local intrinsic clock and the mathematical background clock, and their difference just embodies the curve of spacetime. As for the gravitational redshift effect of light signals emitted from the surface of the sun, strictly speaking, its value should be calculated by incorporating the specific situation of propagations. Since the gravitational field around the sun is in a vacuum spherical symmetry, the metric of spacetime is stationary. In other words, $g_{\mu\nu}$ is irrelevant to the time. Now we assume there are two spatial coordinate points. One is $p_{1}({\bf r}_1)$. Another is $p_{2}({\bf r}_2)$. We introduce a light signal that propagates from $p_{1}$ to $p_{2}$ to investigate the gravitational redshift effect in the solar system. One wavefront is emitted at the moment of coordinate time $t_{1}$ and arrives at $p_{2}$ at the moment of coordinate time $t_{2}$. Thus the time interval measured by the observer's clock (or mathematical background clock) is $\delta t=t_{2}-t_{1}$. Similarly, for the propagation of the next wavefront whose phase difference is $2\pi$, also from $p_{1}$ to $p_{2}$, the time interval measured by the observer's clock is $\delta t'=t'_{2}-t'_{1}$. Considering that the spacetime around the sun is stationary, we have
\begin {eqnarray}
\delta t=\delta t',
\end {eqnarray}
which further indicates
\begin {eqnarray}
dt_{2}\equiv t'_{2}-t_{2}=t'_{1}-t_{1}\equiv dt_{1}.
\end {eqnarray}
Above equation means that the light signal will keep the cycle time and frequency invariant, when it is measured by the observer's clock (or mathematical background clock) in its propagation to any positions in the gravitational field.

As a fundamental assumption, the gravity causes the curve of spacetime but in an infinitesimal neighborhood the spacetime should be asymptotically flat. Therefore, as far as the infinitesimal interval of spacetime is concerned, we are always able to write down the local intrinsic time interval $d\tau$ for every infinitesimal intervals between arbitrary two events: $(t_1,{\bf r}_1)$ and $(t_2,{\bf r}_2)$,
\begin {eqnarray}
ds^2&=&-(1-\frac{2GM}{r})dt^2+(1-\frac{2GM}{r})^{-1}dr^{2}+r^{2}d\theta^{2}+r^{2}sin^{2}\theta d\phi^{2} \cr
&\cong&-d\tau^2(r)++(1-\frac{2GM}{r})^{-1}dr^{2}+r^{2}d\theta^{2}+r^{2}sin^{2}\theta d\phi^{2}.
\end {eqnarray}
For above two wavefronts of the light signals emitted from $p_{1}$ at the moments of $t_{1}$ and $t_{2}$ respectively, it is obvious to have
\begin {eqnarray}
 d\tau_{1}=(1-\frac{2GM}{r_{1}})^{\frac{1}{2}}dt_{1}.
 \end {eqnarray}
Here $\tau_{1}$ is measured by the local intrinsic clock fixed at the spatial coordinate point $p_{1}$, and $t_{1}$ is measured by the observer's clock (or mathematical background clock). Similarly, we have
\begin {eqnarray}
 d\tau_{2}=(1-\frac{2GM}{r_{2}})^{\frac{1}{2}}dt_{2}.
\end {eqnarray}
Therefore,
 \begin {eqnarray}
\frac{d\tau_{1}}{ d\tau_{2}}=\frac{(1-\frac{2GM}{r_{1}})^{\frac{1}{2}}dt_{1}}{(1-\frac{2GM}{r_{2}})^{\frac{1}{2}}dt_{2}}.
 \end {eqnarray}
The frequency measured by the local intrinsic clock satisfies
\begin {eqnarray}
\frac{\nu_{2}}{\nu_{1}}=\frac{d\tau_{1}}{d\tau_{2}}=\frac{(1-\frac{2GM}{r_{1}})^{\frac{1}{2}}dt_{1}}{(1-\frac{2GM}{r_{2}})^{\frac{1}{2}}dt_{2}}.
 \end {eqnarray}
We investigate a following practical case: $p_{1}$ is at rest with respect to the surface of the sun and $p_{2}$ is at rest on the Earth. Since above $d\tau_{1}$ and $d\tau_{2}$ are both corresponding to one cycle time (namely $2\pi$), in consideration of $dt_{2}= dt_{1}$, we also have
 \begin {eqnarray}
\frac{\nu_{2}}{\nu_{1}}=\frac{d\tau_{1}}{ d\tau_{2}}=\frac{(1-\frac{2GM}{r_{1}})^{\frac{1}{2}}}{(1-\frac{2GM}{r_{2}})^{\frac{1}{2}}}<1.
 \end {eqnarray}
Here the frequency of the light signal $\nu_{2}$ is measured by the local intrinsic clock at $p_{2}$. Combining with a fundamental hypothesis that the local frequency of light signal emitted at the surface of the sun is equal to that emitted on the Earth measured by the same kind of local clock on the Earth, then we can draw a conclusion that the frequency of the light signal emitted from the sun is decreased when it is observed on the Earth, compared with the light signal emitted by the same type of atom on the Earth. Ultimately, we demonstrate that the gravitational redshift effect in solar gravity tests can also be self-consistently interpreted by the new proposed physical picture of spacetime.

\section{7 THE VERIFIABLITY OF NEW PHYSICAL PICTURE OF SPACETIME}
In proposed physical picture of spacetime, the most important concept is the absolute background which exists for spacetime. In fact, the background of spacetime can be directly perceived. On the macroscopic scale, any empty space which we have seen is actually a part of the absolute background of space. For example, if an object is taken away from a certain place, the spatial region originally occupied by this object will not disappear with the removing of the object. The existence of this phenomenon partly reflects the existence of an absolute background of space. On the cosmological scale, the background of space is just the common background which reflects the motion of all galaxies in the universe. For instance, when any two adjacent galaxies continually moved away from each other, the empty space vacated between them is a highly approximated background of space. Therefore, the background of space, namely the background of the whole universe, is infinite. There is no concept of volume for the background of space itself. But the commonly referred universe has a size, so it substantially refers to a universe with matter. Here the universe with matter should be conceptually distinguished from the absolute background of space. In this sense, the so-called cosmic accelerated expansion should be more accurately understood as the expansion of the matter inside the absolute background of the universe. Even if there is no matter outside the edge of the current observable universe, we believe that at least the empty space as the background of the universe still exists. Besides, logical argument in Sec.1-3 has also supported that there must be a background of spacetime which exists, as long as the mechanical motion of objects in the universe is real. Therefore, the existence of the background of spacetime is an irrefutable fact.

As for the absoluteness of the background of spacetime and its resulting physical picture of spacetime incorporating relative length or duration of physically defined base units and absolute background, there are still some potential evidences or testable physical features. In this section, we list out three main points as follows.

First, the existence of an absolute background of spacetime is logically testable. In fact, in previous discussion, we have pointed out that the new particle dynamics equation (6) is surly more accurate than a theoretical Newton's second law under the framework of classical mechanics. But the most natural and reasonable derivation of this equation requires nothing but the existence of an absolute background of space. In this sense, the absoluteness of the background of space has been logically proved. Similarly, there might be other physical laws which also potentially support the existence of an absolute background of spacetime if we reinvestigate the physical logic for existing physical theories.

Second, the running rate of a free falling clock under a gravitational field deserves to be further examined. On the one hand, we have logically proved that the nature of the inertial force is the real force acting on the reference object. Hence the so-called inertial force can actually be all kinds of common forces such as friction force, traction force, gravitational force and so on. However, the concept of inertial force still exists in Einstein's special theory of relativity, and even in his general theory of relativity Einstein's equivalence principle still claims that the inertial force is physically equivalent to the gravitational force. But so far as we know, there is only gravitational force has the time dilation effect. On the other hand, whether the redshift effect can be aroused by the acceleration can be tested in a ground-based laboratory, and there has been some high energy experiments showing that the proper longevity of negative muon is not related to its acceleration\cite{Bailey,Farley}. Therefore, whether the free falling clocks under different strength of gravitational fields run at the same rate is totally deserved to be tested. Especially, for the clock which is relatively rest in the gravitational field and the clock which is free falling in the same gravitational field, they differ only in a non-gravitational force and the resulting acceleration. If there is really no gravitational time dilation effect for the free falling clock under changing gravity, it must imply that a non-gravitational force and the resulting acceleration are also able to bring about a time contraction effect for clocks. However, by now there is no such a sign which has been observed and verified in all past experiments. As the priority, a further investigation may be taken on the following point: we may naturally assume that there are two atoms of the same kind exist on the same position of the surface of the sun. Both of them are instantaneously at rest, but one stays at the surface of the sun, and the other starts to be free falling along the radius of the sun at the same moment. The observer on the Earth may detect the light signals emitted by these two atoms and testify whether there is a redshift effect which exists between them.

Third, modern cosmology constitutes the final examination of the absoluteness of the background of spacetime. The existence of an absolute background of spacetime will certainly bring about subtle modifications on the physical picture of Einstein's geometric theory of gravity. An immediate result is that the traditional cosmological metric should be physical amended. A correct cosmological metric should be constructed with fully incorporating at least the following two points. 1, the spacetime should be curved by gravity on the basis of a rigid and homogeneous reference frame of the observer. The clock equipped by the current observer should be imaginarily duplicated onto all moments of the background of time as the standard clock (or mathematical background clock). Similarly, the ruler equipped by the observer on the Earth should be imaginarily duplicated onto all positions of the background of space as the standard ruler (or mathematical background ruler). Then the geometric effect of gravitation can be described by making comparisons between the local physical clock, local physical ruler and above mathematically defined standard clock, standard ruler. Especially, for the cosmology, the present observer on the Earth is the only qualified reference observer to determine all redshift values for all light signals that were emitted from the earlier universe. Therefore, the standard clock and ruler must be defined according to the physical clock and ruler equipped by the observer himself on the Earth at the present time. 2, we know that the matter density in the universe has changed a lot from the beginning of the universe, so the intensity of gravitational field has also changed appreciably. Therefore, if we give up the assumption that all free falling clocks in gravitational fields run in a uniform rate, the intrinsic clock at the present time on the Earth must run at a different rate comparing with that in the earlier universe because of the existence of gravitational time dilation effect. In other words, an evolution of the running rate exists for every local clock fixed on comoving galaxies of the universe. Therefore, with respect to the long evolution history of the universe studied in cosmology, the construction of cosmological metric must exactly distinguish the local intrinsic clock fixed at comoving galaxies and the mathematical clock introduced by the observer at the present time on the Earth (namely coordinate clock). If the reading number of the mathematical clock introduced by the observer at the present time on the Earth is denoted by $t$ and that of the local intrinsic clock fixed at comoving galaxies is denoted by $\tau$, the most general form of cosmological metric under the condition of the cosmological principle is obtained,
\begin {eqnarray}
ds^2=-b^{2}(t)dt^2+a^2(t)[\frac{dr^{2}}{1-kr^{2}}+r^{2}d\theta^{2}+r^{2}sin^{2}\theta d\phi^{2}].
\end {eqnarray}
It should be noticed that we must set $b(t_0)=1$ in the gravitational time dilation effect $d\tau=b(t)dt$, which just means that only at the present time ($t_{0}$) the reading number of the local intrinsic clock fixed at comoving galaxies ($\tau$) reduces to the coordinate time ($t$) which is always imaginarily measured by the intrinsic clock of the present observer on the Earth. We propose the metric (17) to replace the well-known Friedman-Robertson-Walker ($FRW$) metric\cite{weinberg} in processing cosmological observation data. The direct reason is what we reiterated in this paper that cosmological observations are always implemented by the observer at the present time on the Earth, instead of any other observers including the comoving observer in the earlier universe. And the free falling clock is not assumed any more to run at the same rate under evolving gravitation. The new cosmological metric (17) is the one of most important predictions from the absolute background of spacetime and proposed physical picture of spacetime.

\section{8 CONCLUSION}
Starting with natural considerations, we have proposed a fundamental physical picture for spacetime which is compatible with the main physical logic in Einstein's theories of relativity. There are two key points argued in this paper to support a new physical picture of spacetime. The first key point is the introduction of an absolute background of spacetime, meanwhile all previous physical laws about spacetime ( including Einstein's special relativity and general relativity) can deliberately boil down to the evolution law of base units of spacetime. For this proposal, we investigate the formalism of particle dynamics under the non-relativistic framework of classical mechanics, and the nature of the inertial force is revealed by the new particle dynamical equation (6) as the real force acting on the reference object. The second key point is that the running rate of all free falling clocks in gravitational fields is not assumed to run in a uniform rate any more. Therefore, the observation theory for the geometric theory of gravitation is changed. And the physical scenario how gravitation can be converted into a spacetime metric is clarified. To further examine the proposed physical picture of spacetime, we successfully reinterpret the gravitational time dilation effect in solar gravity tests and also point out some possible ways to verify the correctness of our main ideas. In this way, a mutual complementary physical picture of spacetime with relative length or duration of physically defined base units and absolute background is fully presented.

{\bf{Acknowledgments:}}The author would like to thank all his friends who have provided their
constructive criticisms and comments. This work has been supported from the Nature Science Foundation of Zhejiang Province
under the grant number Y6110778. This research was supported in part by the Project of Knowledge Innovation Program (PKIP) of Chinese Academy of Sciences, Grant No. KJCX2.YW.W10. This article is dedicated to my grandfather Mr. Cheng-Dong Chen.

\end{document}